\documentstyle[12pt]{article}
\input{epsf} 

\begin{document}

\begin{center} 
{\bf Predictions for hadron polarizations and left-right asymmetry in inclusive reactions involving photons.}

\bigskip

V. Gupta$^{1}$ and C.J. Solano Salinas$^{2}$\\
CINVESTAV Unidad Merida\\
APDO 73, Cordemex\\
97310-Merida, MEXICO\\
$^{1}$ virendra@mda.cinvestav.mx\\
$^{2}$ javier@mda.cinvestav.mx\\

\medskip

H.S. Mani\\
The Institute of Mathematical Sciences\\
C.I.T. Campos, Chennai 600 113, India\\

\bigskip
February 16, 2004\\

PACS numbers: 13.88.+e, 13.85.Ni, 13.85.Hd, 14.20.Jn
\end{center}

\bigskip
\begin{abstract}
A phenomenological model which has had some success in explaining polarization phenomena and left-right asymmetry in inclusive proton-proton scattering is considered for reactions involving photons.
In particular, the reactions (a) $ \gamma + p \rightarrow H + X;$ (b) $\gamma + p (\uparrow) \rightarrow \pi^{\pm} + X $ and (c)  $p(\uparrow) + p \rightarrow \gamma + X$ are considered where $\gamma =$ resolved photon and hyperon $ H = \Lambda^0, \Sigma^{\pm}$ etc.
Predictions for hyperon polarization in (a) and the asymmetry (in (b) and (c)) provide further tests of this particular model. Feasibility of  observing (b) at HERA and the effect of the polarization of the sea in the proton in $p (\uparrow) + p \rightarrow \pi^{\pm} + X$ is briefly discussed.
\end{abstract}


Polarization phenomena in high energy proton-proton scattering has been studied experimentally~\cite{lesnik} and theoretically~\cite{felix,brqm} for over two decades. Left-right asymmetry $(A_N)$ has been measured~\cite{saroff} in the scattering of a transversely polarized proton $(p({\uparrow}))$ with an unpolarized proton (p) in the inclusive reactions of the type
\begin{equation}
p({\uparrow}) + p {\rightarrow} h + X
\label{eq1}
\end{equation}

where h is a hadron and typically is $\pi^\pm$, $\Lambda$ etc. Significant transverse polarization $P(H)$ for hyperons $H = \Lambda,~ \Sigma^\pm, \Xi^-, \Xi^0, \Sigma^0$ is found in unpolarized proton-proton inclusive scattering, viz,
\begin{equation}
p + p \rightarrow H + X
\end{equation}

The effects observed in the above reactions are non-perturbative by nature and hence have not been accessible to calculations based on first principles, that is, quantum chromodynamics. Instead, phenomenological models have been suggested for the production of the observed hadron.\\

Of the various models, which reproduce some features of the data for $A_N$ and $P (H)$, the most succesful one is perhaps the ``orbiting valence quark model'', sometimes also refered as ``Berliner Relativistic Quark Model (BRQM)''~\cite{brqm} is very interesting in that it relates $A_N$ and $P(H)$ as being due to the same underlying production mechanism in terms of the constituent quarks of the protons and the produced hadron.\\

The basic idea is that a transversely polarized quark $q^P(\uparrow)$  or $q^P(\downarrow)$ in the projectile (P) proton combines with the appropriate quarks or anti-quarks from the `sea' of the target proton to form the observed hadron $h$ in the reaction Eq (1). \\

\begin{figure} 
\centerline{\epsfxsize=6.in \epsffile{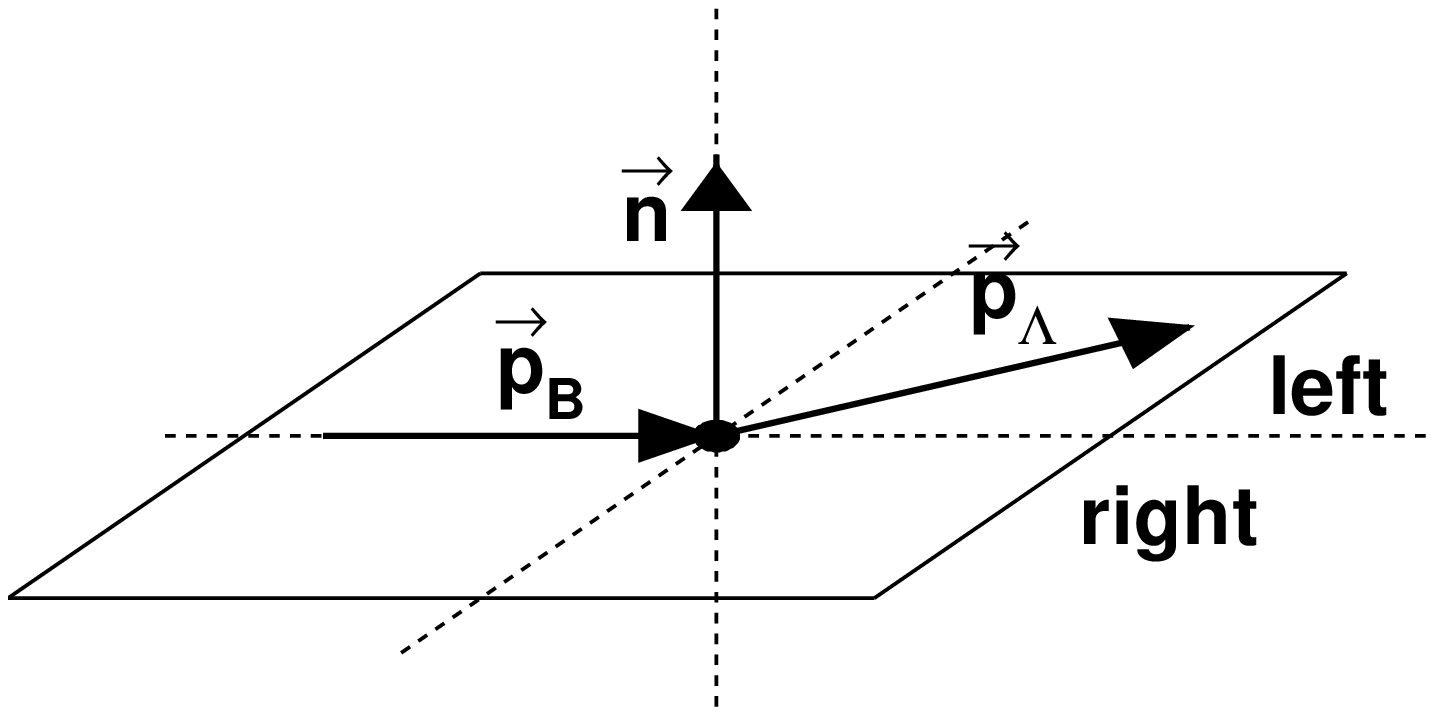}}
\caption{The transverse hadron polarization is defined with respect to the unit vector $\vec n \equiv \frac{\vec p_B \times \vec p_{\Lambda}}{\mid \vec p_B \times \vec p_{\Lambda} \mid}$ which is perpendicular to the production plane, formed by the beam and the produced hadron directions. Here $\vec p_B$ and $\vec p_{\Lambda}$ are momentum of the beam particle and that of the produced hadron $\Lambda$, respectively.}
\label{pol-plane}
\end{figure}

Further, the produced hadron $h$ moves {\underline{preferentially}} to the left of the beam direction, in the upper side of the production plane (see figure~\ref{pol-plane}). That is if $\vec S \cdot \vec n > 0$, where $\vec S$ represents the transverse polarization of $q^P (\uparrow)$, which is polarized upwards w.r.t. the production plane. While, $h$ formed from $q^P(\downarrow)$, with $\vec S \cdot \vec n < 0$,  will move preferentially to the right. This gives rise to a left-right asymmetry. The same reasoning gives rise to a net $P(H)$ in the sub-sample of hyperons going left since H is assumed to retain the polarization of the $q^P$ which forms it~\cite{brqm}.\\

Based on these ideas, we consider reactions (other than in Eqs. (1) and (2)) to provide further tests of this model. In particular, in \S 2 we consider 
(a) the left-right asymmetry in the reaction
\begin{equation}
e^- + p (\uparrow) \rightarrow e^- + \pi^{\pm} + X
\end{equation}
and (b) the transverse polarization of the hyperon H in
\begin{equation}
e^- + p \rightarrow e^- + H + X .
\end{equation}
 In both cases the photon from the $ee\gamma$ vertex is considered to be `real' with `resolved' hadronic components.\\

The left-right asymmetry $A_N^\gamma$ of the photon in 
\begin{equation}
p (\uparrow) + p \rightarrow \gamma + X
\label{eq-prpr-interact}
\end{equation}
 is considered in \S 3. This reaction has a much larger cross-section than the reaction 
\begin{equation}
p (\uparrow) + p \rightarrow e^+ + e^- + X
\end{equation}
 considered earlier [3].\\

In addition, in \S 4 we consider the possibility that the sea of the polarized projectile proton is polarized and estimate the left-right asymmetry of $\pi^\pm$ produced in the target fragmentation region $(x_F < 0)$ in the reaction Eq (1). Lastly, in concluding remarks we summarize our main results.\\


\S 2. {\underline{Resolved photon-proton reactions}}\\

(a) Left-right asymmetry in $e^- + p(\uparrow) \rightarrow e^- + \pi^{\pm}
+ X $ .\\

We consider the photon from the $ ee\gamma$ - vertex with $ M^2 = (mass)^2 \leq (100 MeV)^2$. At such small values of $M^2$ the `resolved' photon $-\gamma$, containing hadronic components dominates in contrast to the `direct' photon which has a point-like interaction. The study of jet production in $e^- + e^+ \rightarrow e^- + e^+ + X$ and $ e^- +p \rightarrow e^- + X$ has led to the determination of the photon structure functions : $ q^{\gamma (M^2)} (x, Q^2)$ for q = u,d,s quarks and $g^{\gamma (M^2)} (x, Q^2)$ for the gluons \cite{drees,grs99}. Here, $Q^2$ is the momentum transfer scale between the $\gamma$ and proton in the effective reaction 
\begin{equation}
\gamma\ + p (\uparrow) \rightarrow  \pi^{\pm} + X
\label{eq-phpr-interact}
\end{equation}

 We now consider the left-right asymmetry of the $\pi^{\pm}$ in this reaction.\\

In the picture under discussion, the $u_v$ valence quark of the proton, with structure function $u^p_v (x^p, Q^2),$ combines with the \={d} in the resolved $\gamma$, with structure function $ \bar{d}^\gamma (x^\gamma, Q^2)$, to form a $\pi^+$ meson. Similarly, $d^p_v$ will combine with the \=u in the photon to form $\pi^-$. We consider the reaction in Eq (7) in the photon-parton center of mass frame with total energy $\sqrt {s}.$ In this frame~\cite{note1} the Feynman parameter $x_F$ and the longitudinal momentum $p_{||}$ for the emitted pion are given by $x_F = x^p - x^{\gamma}$ and $p_{||} = \frac{x_F\sqrt{s}}{2}$, for the proton fragmentation region $x_F > 0$.
The normalized number density of the observed pion in a given kinematic region D is
\begin{equation}
N(x_F, Q | s,i) = \frac{1}{\sigma_m} \int_D d^2p_T \frac{~~d^3\sigma (x_F, \vec{p}_T,Q|s,i)}{dx_Fd^2p_T}
\end{equation}

where $ i = \uparrow$ or $\downarrow$ refers to the transverse spin of the proton, and $\sigma_{in}$ is the total inelastic cross-section.
Then,
\begin{equation}
\Delta N (x_F, Q |s ) = N (x_F, Q|s \uparrow) - N (x_F, Q|s \downarrow) = C_\gamma \Delta D (x_F, Q|s)
\label{eq9}
\end{equation}

where $C_\gamma$ is a flavor independent constant which must be determined fitting experimental data. It is expected to lie between 0 and 1, like the corresponding constant $C \simeq 0.6$~\cite{brqm} for reaction in Eq.~(\ref{eq1}). In Eq.~(\ref{eq9}), $\Delta N$ has been taken to be proportional to $\Delta D (x_F,Q|s) = D (x_F,Q,+|s) - D (x_F,Q,-|s)$ where $D (x_F,Q,\pm |s)$ is the normalized number density eg. of the $(u_v \bar{d})$ which give the $\pi^+$. 

\bigskip

\begin{figure} 
\centerline{\epsfxsize=6.in \epsffile{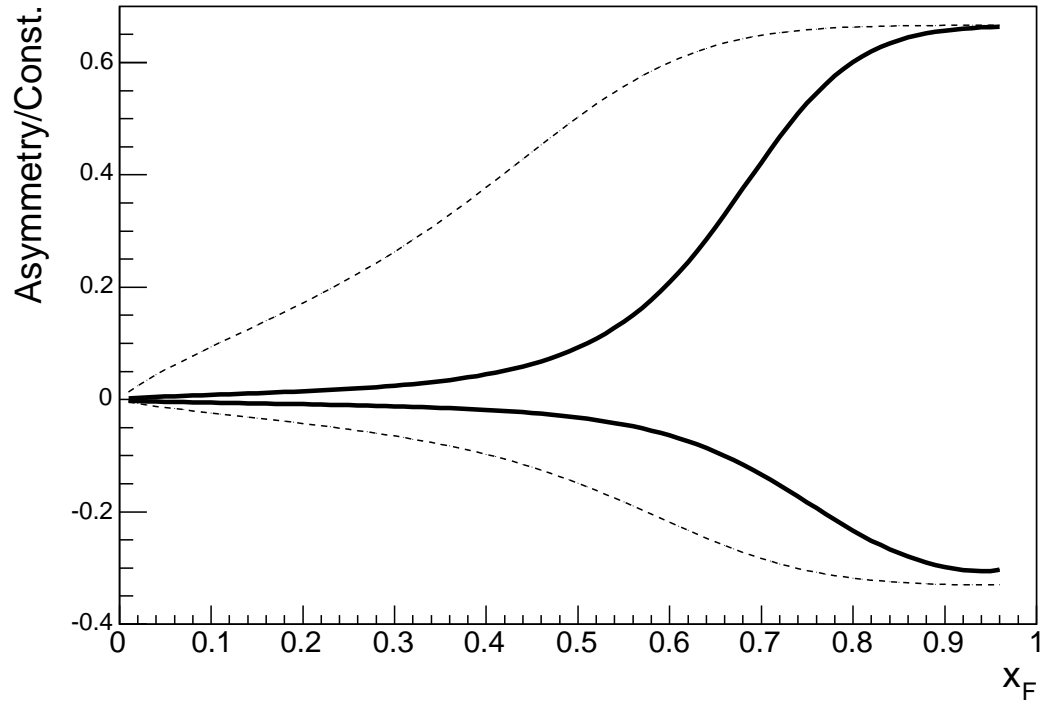}}
\caption{Left-right asymmetry for pions as a function of $x_F$ (See Eq.~(\ref{eq-lrasy-phprpi})) for $\gamma p$ reactions in Eq.~(\ref{eq-phpr-interact}) is plotted as solid lines. The parton distribution GRS99~\cite{grs99} for the photon and CTEQ5~\cite{cteq5} for the proton were used. For comparison, the corresponding quantity for reaction in Eq.~(\ref{eq1}) is plotted using the same parton distributions (dashed lines). In both cases the positive values of asymmetry are for $\pi^+$ and the negative values for $\pi^-$. In the ordinate the constant is $C_{\gamma} ~(C)$ for reactions in Eq.(\ref{eq-phpr-interact}) (Eq.(\ref{eq1})).
}
\label{fig-lrasy-phpr-pi}
\end{figure}

The $\pm$ in $D (x_F,Q,\pm |s)$ refers to the polarization of the valence quark with respect to proton spin direction. We must point out that here, and in the rest of this work, that we are using parton distribution functions which go to zero at $x_F = 1$ and have a finite value for $x_F = 0$. In the model~\cite{brqm}, for $\pi^+$ one has
\begin{equation}
D^{\pi^{+}} (x_F, Q, \pm |s) = K_{\pi}u_v^p  (x^p, Q^2, \pm) \bar{d}^{\gamma} (x^{\gamma}, Q^2)
\end{equation}

where $K_{\pi}$ is a constant. Thus, the asymmetry~\cite{note2}
\begin{equation}
A^{\pi{^+}}_{\gamma_p}(x_F, Q|s) = \frac{C_{\gamma} K_{\pi} [\Delta u^p_v(x^p, Q^2)]\bar{d}^{\gamma}(x^{\gamma}, Q^2)} {N_0(x_F|s) + K_{\pi}u^p_v(x^p,Q^2)\bar{d}^{\gamma}(x^{\gamma},Q^2)}
\label{eq-lrasy-phprpi}
\end{equation}

where the numerator is $\Delta N (x_F, Q|s)$ and Eqs. (9-10) have been used, so that $\Delta u^p_v (x^p, Q^2) = u^p_v (x^p, Q^2, +) - u^p_v (x^p, Q^2, -)$. 
In the denominator, $N_0^{\gamma} (x_F |s)$ stands for the non-direct part of $\pi^+$ production and the second term is $[D^{\pi^+} (x_F, Q, + |s) + D^{\pi^+} (x_F,Q, -|s)]$.
In general, the non-direct part $N_0^{\gamma}$ for reaction in Eq.~(\ref{eq-phpr-interact}) and the non-direct part $N_0$ for reaction in Eq.~(\ref{eq1}) will be different.However, one expects that their behaviour with respect to $x_F$ will be similar. From fits to experimental data in \cite{boros}, for $h=\pi^{\pm}$ in Eq.~(\ref{eq1}), $N_0 (x_F|s)$ is known. Assuming $N_0^{\gamma} (x_F|s) \simeq N_0 (x_F|s)$, the asymmetry in Eq.~(\ref{eq-lrasy-phprpi}) is plotted in Fig.~\ref{fig-lrasy-phpr-pi} using the experimentally known $\bar d^{\gamma}$ and $u_v^p$~\cite{grs99,cteq5}.
The non-direct part $N_0 (x_F|s)$ is significant for small $x_F$ and negligible for $x_F > 0.5$. One expects $N_0^{\gamma} (x_F|s)$ also to be negligible for large $x_F$.\\

In Fig.~\ref{fig-lrasy-phpr-pi} we can see that results for photon--proton reaction are qualitatively similar to those for pion production in proton--proton reactions. the latter are recalculated using the new proton distribution functions CTEQ5~\cite{cteq5} and show asymmetry for lower $x_F$ values than before~\cite{boros}. Note that the asymmetry in both cases is qualitatively the same for $x_F > 0.8$.\\

If one neglects the non-direct contribution for $x_F > 0.5$, Eq.~(\ref{eq-lrasy-phprpi}) simplifies to
\begin{equation}
A^{\pi^+}_{\gamma p} \simeq C_\gamma  \frac{\Delta u^p_v (x^p,Q^2)}{u^p_v (x^p, Q^2)} \simeq \frac{2}{3} C_\gamma
\end{equation}
The second equality is obtained using $SU(6)$ wave function for the proton since one has $\bigtriangleup u_v(x) = \frac{2}{3} \ u_v(x)$.
This predicts that for large $x_F (> 0.5)$ the asymmetry $A^{\pi^+}_{\gamma p} / C_{\gamma}$ is positive and similar to $A^{\pi^+}_{pp} / C$ seen in $p(\uparrow) + p \rightarrow \pi^+ + X$. Further, numerically the value for large $x_F$ is $\simeq \frac{2}{3}$ (for both cases), as seen in Fig.~\ref{fig-lrasy-phpr-pi}.

\bigskip

Using analogous arguments for $\pi^-$ one predicts $A^{\pi^-}_{\gamma p}$ to be smaller (than $A^{\pi^+}_{\gamma p}$) but negative and similar in behavior to $A^{\pi^-}_{N}$. Again, neglecting the non-direct contribution for large $x^\gamma_p > 0.5$ and using $SU(6)$ wave function (i.e. $\bigtriangleup d_v(x) = -\frac{1}{3} \ d_v(x)$) we obtain $A^{\pi^-}_{\gamma p} \simeq - \frac{1}{3} C_\gamma$. We discuss the feasibility of observing this asymmetry experimentally in $\S 2c$.

\bigskip

(b)  Transverse polarization of the hyperon $(H)$ in $e^- + p
\rightarrow  e^- + H + X$.

\bigskip

Here the effective reaction is
\begin{equation}
\gamma + p \longrightarrow H + X
\label{eq-phpr-h}
\end{equation}

where the $p$ is unpolarized and $ H = \Lambda, \Sigma^\pm$ etc.

\bigskip

The kinematics is similar to that for reaction in Eq (7), except that here the hyperon polarization P(H) is measured. To see how P(H) arises in the model, take the direction of the proton beam in cm-frame (see figure~\ref{pol-plane}). Let transverse polarization of a quark (in the proton) be denoted by $\uparrow$ or $\downarrow$, perpendicular to the production plane.
One assumes~\cite{brqm}, that a quark with upward ($\uparrow$) (downward ($\downarrow$)) polarization will preferentially scatter to the left (right) in the production plane, with respect to the beam direction.

\bigskip

This quark will combine with a two quark state $(qq)_{\gamma}$ from the photon to form the hyperon  H. It is also possible that two quarks from the proton combine with a quark from the photon to give the polarized H. To clarify how the model works let us consider the production of $\Sigma^-$ whose quark content is $dds$. In this case, only the valence d-quark from the proton ($d^p_v$) is common with those in $\Sigma^-$. Let the probability of $d^p_v(\uparrow) (d^p_v(\downarrow))$ from the proton to move to the left (right) be $\alpha$. Then, the probability of $d^p_v(\uparrow) (d^p_v(\downarrow))$ to move to the right (left) will be ($1-\alpha$). The unpolarized proton has equal probability of having a $d^p_v(\uparrow)$ or $d^p_v(\downarrow)$. Using $SU(6)$ baryon wave functions, we know that the probability of $d\uparrow (d\downarrow)$ in a $\Sigma^-(\uparrow)$ is $\frac{5}{6}$ $(\frac{1}{6})$. We expect that $N(\Sigma^-(\uparrow))$, the number of $\Sigma^-(\uparrow)$ formed by the left moving d$\uparrow$ and $d\downarrow$, will be proportional to $\frac{5}{6} \ \alpha + \frac{1}{6} (1-\alpha)$, while the number $N(\Sigma^-(\downarrow))$ will be proportional to $\frac{1}{6} \alpha + \frac{5}{6} (1-\alpha)$.  

\bigskip

Thus, one expects that the polarization 
\begin{equation}
P(\Sigma^-) = \frac{N(\Sigma^-(\uparrow)) - 
N(\Sigma^-(\downarrow))}{N(\Sigma^-(\uparrow)) +
N(\Sigma^-(\downarrow))} = \frac{2}{3} \ (2\alpha - 1)
\end{equation}

As expected, this is zero if $\alpha = \frac{1}{2}$. Since, $\alpha$ is assumed to be $\geq 1/2$, the model predicts $0\leq P(\Sigma^-) \leq 2/3$.
Using $SU(6)$ wave functions, we analyze the expected polarization for the other hyperons.
\begin{enumerate}
\item[(i)] For $\Xi^0(uss)$ and $\Xi^-(dss)$, the $u^p_v$ and $d^p_v$ will contribute to their production by combining with the $(ss)_{\gamma}-$ state from the photon. 
Here $N(\Xi^{0\uparrow})$ or $N(\Xi^{-\uparrow})$ 
is $\propto[1/3 \ \alpha + \frac{2}{3} \ (1-\alpha)]$, 
while $N(\Xi^{0\downarrow})$  or  $N(\Xi^{-\downarrow})$ 
is $\propto[\frac{2}{3} \ \alpha + \frac{1}{3} \ (1-\alpha)]$, 
so that
\begin{equation}
P(\Xi^0) = P(\Xi^-) = \frac{1}{3} \ (1-2\alpha).
\end{equation}

The polarization is expected to be opposite in sign to $P(\Sigma^-)$ but
smaller by  factor of two.

\item[(ii)] For $\Sigma^+ (uus)$ there are two formation mechanisms : $(1) u^p_v + (us)_\gamma$ and $(2) (uu)^p_v + (s)_\gamma$. The first mechanism (as in the $\Sigma^- $ case) will contribute $\frac{2}{3} \ (2\alpha - 1)$ to the polarization. If $\alpha_2$ is the probability the spin up diquark (uu) to move to the left then the second mechanism will contribute $\frac{2}{3}(2\alpha_2-1)$. If ${\mathcal{P}}_1$ and ${\mathcal{P}}_2$ are the probabilities for $\Sigma^+$ to be formed by the mechanisms $(1)$ and $(2)$ respectively then, one expects to have
\begin{equation}
P(\Sigma^+) = \frac{2}{3} \ (2\alpha - 1){\mathcal{P}}_1 + \frac{2}{3} (2\alpha_2 - 1){\mathcal{P}}_2
\end{equation}

There is no definite prediction for the polarization without having a way to estimate ${\mathcal{P}}_1, {\mathcal{P}}_2$ and $\alpha_2$. Of course, if mechanism $(1)$ dominates (i.e. $P_2\simeq 0$) then one expects $P(\Sigma^+)\simeq P(\Sigma^-$) or smaller.

\item[(iii)] For $\Sigma^0 (uds)$ there are 3 formation mechanisms: $(1) u^p_v + (ds)_\gamma, (2) d^p_v + (us)_\gamma,$ and $(3) (ud)^p_v + (s)_\gamma$. First two are single valence quark formation as in $\Sigma^\pm$ while the third is like the diquark mechanism in $\Sigma^+$. If ${\mathcal{P}}_i ~(i= 1,2,3)$ are the the probabilities for these three mechanisms, then
\begin{equation}
P(\Sigma^0) = \frac{2}{3} (2\alpha - 1)({\mathcal{P}}_1 + {\mathcal{P}}_2) + \frac{2}{3} (2\alpha_2 - 1) {\mathcal{P}}_3
\end{equation}

Again, if the valence diquark mechanism is negligible then one may expect $P(\Sigma^0)$ to be similar to $P(\Sigma^-)$.

\item[(iv)] For $\Lambda^0(uds)$ the situation is different since the polarization of $\Lambda^0$ comes only from the s-quark. Here the mechanism suggested ${[3]}$ is that $\Lambda^0$ is produced in association with a $K^+ (u^p_v\bar{s}_\gamma)$ or $K^0 (d^p_v\bar{s}_\gamma)$ which takes the $\bar{s}_\gamma$ for the spin zero pair $(s\bar{s})_\gamma$ from the photon. 
So, when $u^p_v\uparrow$ turns preferentially to the left with probability $\alpha$ it forms a $K^+$ with $\bar{s} _\gamma\downarrow$ leaving $s_\gamma\uparrow$ to form a $\Lambda^0\uparrow$ state moving to the right. 
So
\begin{equation}
P(\Lambda^0) = - (2\alpha - 1),
\end{equation}

which is opposite in sign to $P(\Sigma^-)$ but larger in magnitude. For $\alpha > 1/2$ $P(\Sigma^-)$ is positive while $P(\Lambda^0) = 3P(\Xi^0) = 3P(\Xi^-)$ are negative as in case of reaction in Eq(2).
\end{enumerate}

\bigskip

However, one should bear in mind the possibility (not considered in~\cite{boros}) that the $\Lambda^0$ can also be produced in association with a $K^{*+}$ or a $K^{*0}$. This would modify Eq (18) to 
\begin{center}
$ P(\Lambda^0)  = (2 \alpha - 1) (1 - 2 \eta) ~~~~~~~~~~~~~~~~~~(18a)$ 
\end{center}

where $\eta$ and $(1-\eta)$ denote the probabilities for the $\Lambda^0 K$  or $\Lambda^0 K^*$ channels. 
For $\eta =1$ (no $\Lambda^0K^*$ channel) we recover Eq (18). The key point
is that if $\eta < \frac{1}{2}$ then, for $\alpha > \frac{1}{2}$, $P(\Lambda^0)$ will be positive. Note that the value of $\eta$ will depend on the reaction. It could be $\approx 1$ for the reaction in Eq (2) while it could be $< \frac{1}{2}$ in Eq (13). So that $P(\Lambda^0)$ would be negative for the first reaction but positive for the latter reaction.\\

\noindent(c)  We briefly consider the experimental possibilities for studying the reaction in Eq (7).\\


If $E_\gamma$ is the energy of the photon in the frame (the lab-frame) in which the proton is at rest, then $S \simeq 2m_pE_\gamma$. Let $\Theta_L$ and $\Theta_{cm}$ be the angles made by the pion with respect to $-\overrightarrow{p}_\gamma$ in the lab-frame and the cm-frame of $\gamma-p$ respectively.
Here ${p}_\gamma$ is the lab-momentum of the photon. The two angles are related by $(m^2_\pi\simeq 0)$
\begin{equation}
\cos^2\left(\frac{\Theta_L}{2}\right) = \frac{m_p}{2E_\gamma}
\cot^2\left(\frac{\Theta_{cm}}{2}\right)
\end{equation}

For $E_\gamma \simeq 10GeV$ and $\Theta_{cm} \sim 30^o$ this gives a
large value $\Theta_L \sim 70^o$. Since~\cite{note2}, $x^px^\gamma \simeq m^2_\pi/S \simeq 5\times 10^{-4}$, such experiments will probe the structure of the photon in the regions $x^\gamma \simeq 10^{-3}$ for $x^p \sim 0.5$.\\


In HERA, one expects collision of `real' photons of energy $E_\gamma  \sim 20GeV$ with protons of energy $E_p = 820GeV$ giving $s_{\gamma p} \simeq 6.6\times 10^4(GeV)^2$ corresponding to a photon with cm-energy $E^{cm}_\gamma \simeq 128GeV$.\\


If $\Theta _H$ is the angle of the pion in the HERA-frame then one obtains
\begin{equation}
\cot^2\left(\frac{\Theta_H}{2}\right) = \frac{E^H_p}{E^H_\gamma} \cot^2\left(\frac{\Theta_{cm}}{2}\right)
\end{equation}

In the present arrangement in $H1$~\cite{note3} one needs $\Theta_H \sim 30^o$ for the pion to be detectable. Therefore Eq(20) gives an angle of about $\Theta_{cm} \sim 129^o$. HERA experiments would probe regions for which $x^px^r\sim 3\times 10^{-7}$. One can hope to probe rather low values of x in the proton where possible polarization of the sea could well be tested.\\


\vspace{0.5cm}

\S 3. {\underline{Left-right asymmetry in $p(\uparrow) + p\rightarrow \gamma + X$}}.

\bigskip

\begin{figure} 
\centerline{\epsfxsize=6.0in \epsffile{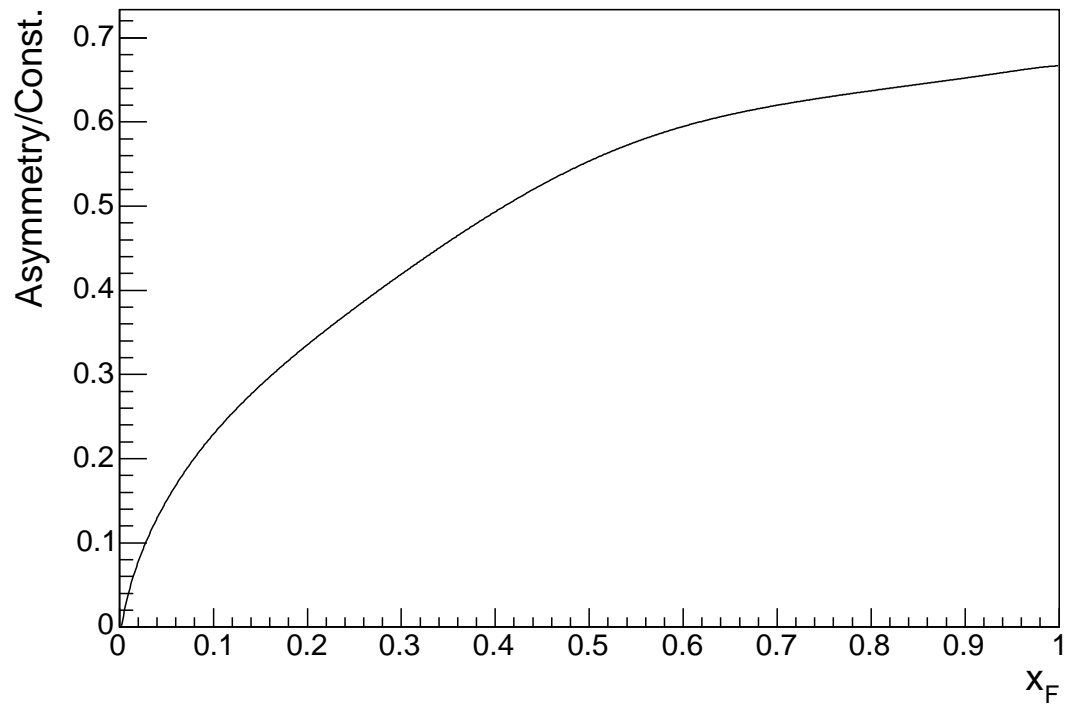}}
\caption{Left-right asymmetry for photons as a function of $x_F$ in proton--proton interactions (Eq.\ref{eq-prpr-interact}). For the non-direct formation we used $N_0^{pp \rightarrow \pi}$~\cite{boros}. The ordinate is $A^\gamma_N(x^\gamma_F)/\tilde{C}_\gamma$ given in Eq.~(\ref{eq-lrasy-prprph}).}
\label{fig-lrasy-prpr-ph}
\end{figure}

In this reaction, if the projectile proton is polarized then one would expect to observe $A^\gamma_N$ the left-right asymmetry of the emitted photon in the projectile fragmentation region. The process at the quark level is ${q}_v + \bar{q}_s \rightarrow \gamma + gluon$, where $q_v = u$ or $d$ and $\bar{q}_s$ is from the `sea' of the other proton. Formation of $\gamma$ through $u\bar{u}$ is larger than through $d\bar{d}$ by a factor $4$. 
For a polarized projectile proton, the model gives the asymmetry
\begin{equation}
A^\gamma_N(x^\gamma_F) = \frac{\tilde{C}_\gamma K_\gamma  [4\bigtriangleup u^P_v(x^P)\bar{u}^T_s (x^T) + \bigtriangleup d^P_v(x^P)\bar{d}^T_s(x^T)]}{N_0(x^\gamma_F)+K_\gamma [4 u^P_v(x^P)\bar{u}^T_s (x^T) +  d^P_v(x^P)\bar{d}^T_s(x^T)]}
\label{eq-lrasy-prprph}
\end{equation}

where the superscripts $P$ and $T$ refer to the projectile and the target and $\tilde{C}_\gamma$ and $K_\gamma$ are constants to be determined. The expected asymmetry for this case is plotted in figure~\ref{fig-lrasy-prpr-ph} where we can see that in comparison to pion production (Fig.~\ref{fig-lrasy-phpr-pi}) the asymmetry becomes important even for small $x_F$ values.\\


Again, assuming $SU(6)$ wave function for the proton, aproximating $\bar{u}_s(x) = \bar{d}_s(x)$ and neglecting the non-direct contribution $N_0$ for large $x^\gamma_p > 0.5$, Eq.(\ref{eq-lrasy-prprph}) simplifies to

\begin{equation}
A^\gamma_N(x^\gamma_F) \simeq \frac{\tilde{C}_\gamma}{3} ~ \frac{8 u^P_v(x^P) - d^P_v(x^P)}{4 u^P_v(x^P) + d^P_v(x^P)}
\end{equation}


Since, for $x_F > 0.8$, $u_v^P$ dominates over $d_v^P$ one expects $A^\gamma_N(x^\gamma_F)/C_{\gamma} \simeq 2/3$. This feature is seen in Fig.~\ref{fig-lrasy-prpr-ph}. It may be noted that $x^T x^P$ is not restricted to be small here since $u\bar{u}$ and $d\bar{d}$ produce a $\gamma + gluon$. Thus, $x^p \simeq x^\gamma_F$ and $x^T~ {^{^<}\!\!\!\!\!\!\sim} ~0.1$ would be a good region to study.\\

\newpage

$\S 4$ \qquad {\underline{Effect of sea polarization}}:

\bigskip

The experimental study {\cite{ashman}} of the $g^p_1$ structure function has led to an interesting possibility that the sea in the proton is strongly polarized. This could lead to a possible observable asymmetry in the target fragmentation region $(x_F<0)$ in addition to that observed in the projectile fragmentation region $(x_F>0)$.\\


Consider the reaction $p(\uparrow) + p \rightarrow \pi^\pm + X$ where the projectile is polarized. In the model under consideration one would expect 
\begin{equation}
{\mathcal{A}}^{\pi +}_N (-|x_F|,s) = \frac{C_s K_\pi \bigtriangleup \bar{d}_s(x^P) u_v(x^T)}{N_0^{pp \rightarrow \pi}(-|x_F|) + K_\pi \bar{d}_s(x^P)u_v(x^T)}
\label{eq-seaprpr-pi}
\end{equation}
where $C_s$ (subscript `s' is for sea) is a constant to be determined. The expression for ${\mathcal{A}}^{\pi -}_N$ can be obtained by $u\leftrightarrow d$ in Eq (23). Note also that $-x^T + x^P = -|x_F|$ and $|x^Px^T| = m^2_\pi/s$.\\


Measurement of such asymmetries would probe the polarization of $\bar{d}_s$ or $\bar{u}_s$ in the sea.\\

One can also probe these by studying $\gamma + p(\uparrow) \rightarrow \pi^\pm + X$. In this case, for $cm.$  energy $\sqrt{s}/2$,
\begin{equation}
{\mathcal{A}}^{\pi -}_{\gamma p}(-|x_F|,s) = \frac{C_s^{\gamma} K_\pi \bigtriangleup \bar{u}^p_s(x^P) d_\gamma (x^\gamma)}{N_0^{pp \rightarrow \pi}(-|x_F|) + K_\pi \bar{u}^p_s(x^P)d_\gamma (x^\gamma)}
\label{eq-seaphpr-pi}
\end{equation}
where $x^P - x^\gamma = -|x_F|$ and $|x^Px^\gamma| = m^2_\pi / s$. 
It is important to note that the asymmetry in Eqs.~\ref{eq-seaprpr-pi} and \ref{eq-seaphpr-pi} are observable for $x_F < 0$ (or $-|x_F|$) in contrast to those discussed in $\S 3$ which are in the region $x_F > 0$. ${\mathcal{A}}_{\gamma p}^{\pi^+}$ can be obtained by $d \leftrightarrow u$ in Eq.~(\ref{eq-seaphpr-pi}.\\

\begin{figure} 
\centerline{\epsfxsize=6.0in \epsffile{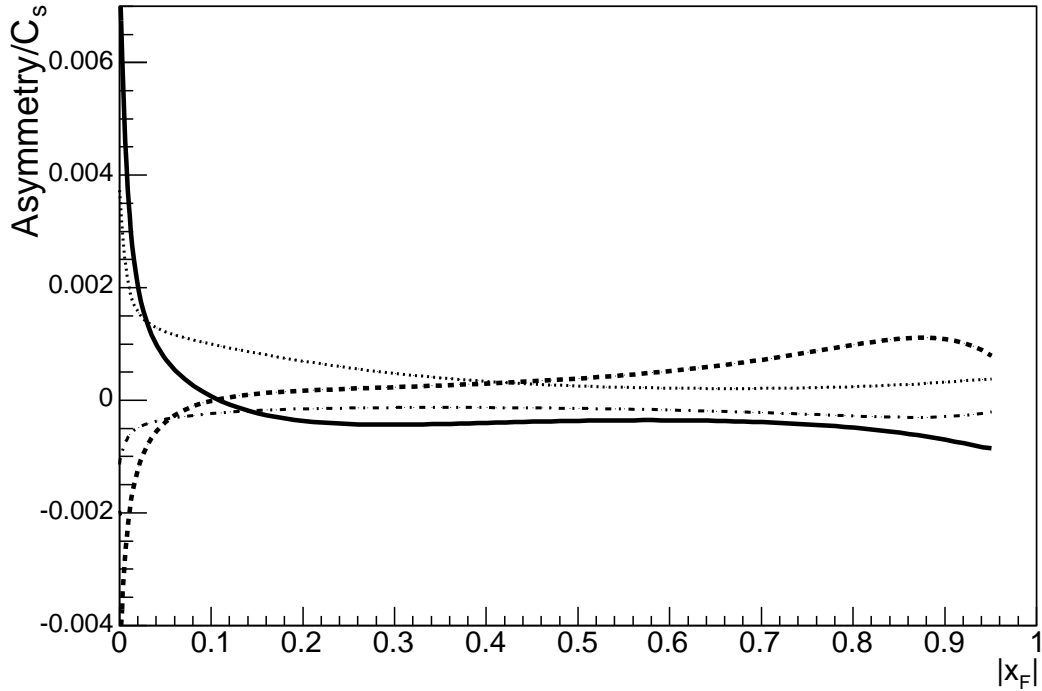}}
\caption{Left-right asymmetry for pions, from sea polarization, as a function of $|x_F|$ for negative $x_F$. In each case the ordinate is the asymmetry (see Eqs.(~\ref{eq-seaprpr-pi}) and (\ref{eq-seaphpr-pi})) divided by the unknown constant $C_s$, to be determined by fitting experimental data. For the proton--proton reactions (in Eq.~(\ref{eq1})), ${\mathcal{A}}_N^{\pi^+} / C_s$ and ${\mathcal{A}}_N^{\pi^-} / C_s$ are given by ``solid'' and ``dashed'' lines, respectively. For photon--proton reactions (see Eq.~(\ref{eq-phpr-interact})) the corresponding quantities ${\mathcal{A}}_{\gamma p}^{\pi^+} / C_s$ and ${\mathcal{A}}_{\gamma p}^{\pi^-} / C_s$ are given by ``dotted'' and ``dot-dashed'' lines.
}
\label{fig-lrasy-phpr-xpi}
\end{figure}

Estimates of $\bigtriangleup\bar{u}_s$ have been recently extracted from the data~\cite{goshtasbpour} giving $\bigtriangleup\bar{u}_s \sim (-0.1)u_s$. The $x_F$ dependence of the asymmetries (divided by the undetermined overall constant $C_s$) in Eqs.~\ref{eq-seaprpr-pi} and \ref{eq-seaphpr-pi} are plotted in Fig.~\ref{fig-lrasy-phpr-xpi}. For the non-direct contribution we used $N_0^{pp \rightarrow \pi}$ given in reference \cite{boros}.\\

From Fig.~\ref{fig-lrasy-phpr-xpi}, we note the following points:\\

a) The asymmetries from the sea polarization are generally very small. for example less tan 0.002 for $x_F < -0.1$ (or $|x_F| > 0.1$). This is in contrast to the asymmetries arising from valence quarks, see Fig.~\ref{fig-lrasy-phpr-pi}.

b) Sea polarization asymmetries are larger near $x_F \simeq 0$ than the valence quark polarization asymmetries.\\

These two points are true for both proton--proton and photon--proton reactions. An interesting feature is to be seen in  Fig.~\ref{fig-lrasy-phpr-xpi} for proton--proton reactions, namely, a cross-over point at $|x_F| \simeq 0.1$, where ${\mathcal{A}}_N^{\pi^+} = {\mathcal{A}}_N^{\pi^-} \simeq 0$.\\




\noindent{\bf Concluding remarks}:

\bigskip

In this paper, tests of a particular phenomenological model are given for some new processes in sections 2-4. In particular, relations like $P(\Sigma^-) = -2P(\Xi^0$) for the process in Eq.~(\ref{eq-phpr-h}) and $A_{\gamma p}^{\pi^+}  \sim \frac{2}{3}~ C_\gamma$ (for $x_F > 0.8$) in $\S 2$, provide new and simple tests of the model. It is interesting to note that $A_N^{\pi^+}  \sim \frac{2}{3}~ C$ (Fig.~\ref{fig-lrasy-phpr-pi}) and $A_N^{\pi^+}  \sim \frac{2}{3}~ \tilde C_\gamma$ (Fig~\ref{fig-lrasy-prpr-ph}) have similar limits.
The processes which provide these tests would probe small x-region of the proton.\\

Further, the possibility of a left-right asymmetry in the target fragmentation region due to a polarized sea is suggested. In all the cases considered here, these asymmetries are small (Fig.~\ref{fig-lrasy-phpr-xpi}).
 We expect that with the new generations of experiments in HERA, and maybe at Jefferson Lab, will be able to measure the effects discussed above.
\bigskip

{\bf\underline{Acknowledgment}} \\

We are grateful to G. Contreras and Julian F\'{e}lix for comments and discussions. One of us HSM is grateful to CINVESTAV Unidad Merida for partial support during his visit there. The work is also supported in  part by Conacyt-Mexico, Project No. 28265E.

\newpage

\newpage

\newpage

\newpage

\begin{thebibliography}{99}
\bibitem{note1} Note $s = (p^p + p^{\gamma})^2 \simeq 2 p^p.p^{\gamma}$, $~Q^2 = (x^p p^p + x^{\gamma} p^{\gamma})^2 \simeq s x^p x^{\gamma}$ and 
$(x^{p} {\vec p}^{~p}_{||} + x^{\gamma} {\vec p}^{~p}_{||}) = {\vec p}^{~\pi}_{||}$. Last equation gives $x^p - x^{\gamma} = x_F$ in the CM-frame. Superscripts $p$ and ${\gamma}$ refer to the proton and the photon while $x^p$ and $x^{\gamma}$ are the fraction of the momentum carried by their constituent involved in the reaction.
\bibitem{note2} Since here $Q^2 = m^2_\pi$, one has $x^p x^{\gamma} \simeq m^2_\pi/s$ which means that small values of $x^{\gamma}$ will be probed.
\end{thebibliography}
\end{document}